\documentclass[twocolumn,fleqn,natbib]{svjour3}
\bibpunct{[}{]}{;}{n}{}{,} % to get "[numbered]" references from natbib
\smartqed  % flush right qed marks, e.g. at end of proof

\usepackage{mathptmx}      % use Times fonts if available on your TeX system
\usepackage{graphicx}

\usepackage{amsmath}

\usepackage{amssymb}

\usepackage{natbib}

\journalname{Granular Matter}

\begin{document}

\title{Stress response inside perturbed particle assemblies}

\author{Leonardo E.~Silbert}

\institute 
{ 
  Department of Physics, Southern Illinois University, Carbondale, IL
  62901, USA,
  \\\email{lsilbert@physics.siu.edu}
}

\date{Submitted;}

\maketitle

\begin{abstract}
  
  The effect of structural disorder on the stress response inside three
  dimensional particle assemblies is studied using computer simulations of
  frictionless sphere packings. Upon applying a localised, perturbative force
  within the packings, the resulting {\it Green's} function response is mapped
  inside the different assemblies, thus providing an explicit view as to how
  the imposed perturbation is transmitted through the packing. In weakly
  disordered arrays, the resulting transmission of forces is of the
  double-peak variety, but with peak widths scaling linearly with distance
  from the source of the perturbation. This behaviour is consistent with an
  anisotropic elasticity response profile. Increasing the disorder distorts
  the response function until a single-peak response is obtained for fully
  disordered packings consistent with an isotropic description.


 
\end{abstract}
 
\section{Introduction}
The internal stresses of a particulate medium, such as a sandpile,
determine mechanical stability and structural robustness. How these
stresses are distributed can have a profound effect on the way such a
system responds to external loading. This is a challenging problem
which, one hopes, will lead to a better understanding of the
transition between quiescence and catastrophic failure, such as a
collapsing grain silo or the onset of an avalanche. 

Over the past decade or so, it has come to light that the way forces are
transmitted through a disordered pile of grains occurs in a rather
inhomogeneous manner \cite{nagel2}. It seems that it is very much the
particulate nature of a granular material that lends itself to the emergence
of strongly heterogeneous force networks in disordered packings.  Consequently
the nature of stress transmission in response to external loading might also
exhibit unusual properties due to granularity. These issues have led to
renewed interest in describing how stresses are transmitted through
particulate matter.

Whereas classical, isotropic, elasticity theory \cite{landau2,johnson1}
accurately describes the stress state of continuous media, when the discrete
nature of the constituent particles influence the way a system responds, it is
unclear how one should proceed. So much so that recent simulations of
Lennard-Jones glasses have shown how the discrete nature of the amorphous
glass influences the elastic properties of the material
\cite{wittmer1,wittmer2,wittmer3,wittmer4}. Thus, in the context of jammed
particulate media, the humble sandpile has thus come to represent the paradigm
of an unusual state of matter exhibiting heterogeneous force distributions.

One of the simplest methods to access information about the stress state of a
material is to measure the Green's function response to a localised
perturbation. For a granular packing, this would involve applying a small
force to the central particle at the top of the packing then measuring the
pressure at the bottom of the packing in response to this localised external
perturbation. To remain in the linear regime, the applied force perturbation
must be sufficiently small so as to avoid rearrangement of the grains from
their original positions. This is the approach taken in several experimental
and simulation studies to date.

Some of the more striking studies involved the direct visualisation of the
force networks in two dimensional ($2D$) packings using photoelastic particles
\cite{behringer4,behringer5,rajchenbach2}, and their micro-displacements
\cite{bonamy1,moukarzel3}. Direct imaging and the subsequent analyses allows
one to determine how the contact and/or force network responds when a point
force is applied to the packing. Presently, more indirect methods are required
for inferring the way forces are transmitted through $3D$ packings
\cite{clement1,nagel6,nagel7,qi1}, simply due to the fact that granular packs
are inconveniently opaque. However, there is a clear distinction between the
response properties of disordered grain piles \cite{clement1} and ordered
arrays \cite{nagel6}. There are also numerous analytical and numerical works
\cite{coppersmith1,clement2,cates3,hurd1,claudin1,claudin8,witten3,goldenberg1,goldenberg2,dasilveira3,blumenfeld2,panja1,claudin6,goldenberg3,makse4,tordesillas1},
predicting how forces are transmitted in response to such perturbations.

As a pr\'{e}cis to the above, also see Ref.~\cite{goldenberg4}, it is now
appreciated that granular materials have the uncanny ability to respond in a
variety of ways. Isotropic, {\it single-peak} response profiles mimic the
manner in which an elastic medium behaves as described by partial differential
equations of the elliptic class. The isotropic elastic stress state, $\sigma$,
of a material in response to a localised load, $F$, on an infinite half space
is described by the Boussinesq equation \cite{johnson1},
\begin{equation}
\centering
\sigma = \frac{3F}{2\pi}\frac{z^{3}}{\left(z^{2}+r^{2}\right)^{5/2}},
\label{eqn1}
\end{equation}
where $z$ is the vertical distance, or depth, from the source of the force
perturbation, and $r$ is the planar distance from the axis of $F$
\cite{landau2,johnson1}. Most of the experiments and simulations performed to
date are constructed in such a manner as to directly test the validity of this
theory and its $2D$ analogue the Flamant equation. Under some conditions,
strongly anisotropic, {\it double-peak} response profiles have been reported.
These types of response profiles can be indicative of anisotropic, elastic
behaviour, and therefore formally belong to an elliptic description. In the
extreme case, double-peak behaviour is representative of hyperbolic,
\emph{wave-like} propagative models, whereby the stresses \emph{propagate}
along characteristic pathways analogous to light rays. In between, is a
\emph{diffusive} parabolic model that seems to be losing favour despite
playing a seminal role in describing force heterogeneities in granular media.
Moreover, a crossover between different response functions can occur, even
within the same pile, depending on structure, particle friction coefficient,
distance from the source of the perturbation, and the magnitude of the imposed
force.  Typically, however, ordered grain configurations result in anisotropic
response features, whereas disordered systems usually conform to an isotropic
picture.

As far as $2D$ results go, many of the features described above can be
accommodated within the framework of $2D$ anisotropic elasticity
theory \cite{goldenberg5,claudin4}. In fact, a parameter space of anisotropic
elasticity theory was proposed \cite{claudin4} that allows for the
appearance of either single-peak or double-peak response functions
within the same formalism. Simply put, anisotropy in the force/contact
network promotes an anisotropic response. One of the key concerns,
therefore, is the ability to distinguish between the elastic and
wave-like response profiles.

Clearly there are many factors that influence the nature of the resulting
response. It would seem prudent to investigate the underlying factors that
have the greatest influence in determining the response properties. This is
precisely the approach taken in this study. Structure, or the arrangement of
the particles that constitute the packing, plays a dominant role. The results
presented here are the first to investigate the manner in which the stresses
{\it inside} $3D$ particle packings are transmitted in response to localised
force perturbations, and how varying the disorder affects the response.

\section{Simulation Model}
The computer experiments reported here are designed to systematically study
how structure influences stress transmission in response to localised
perturbations. To that end, the simulations are carried out on a model system:
three dimensional, monodisperse, non-cohesive, frictionless, soft-sphere
packings, with fully periodic boundary conditions, in the absence of gravity.
Generation of the initial packings has been described in detail elsewhere
\cite{leo16,leo18}. To summarise, $N$ spheres of diameter $d$ and mass $m$,
were arranged into a face-centred cubic (fcc) array. The packing fraction,
$\phi$, of the particle assemblies was fixed at $\phi = 0.742$, just above
that of a hard-sphere fcc array, $\phi_{\rm{fcc}} = \sqrt{2}\pi/6$.  Two
particles, $i$ and $j$, are defined to be contact neighbours and interact
through a short-range, purely repulsive, force when $r_{ij} < d$, where
$r_{ij}$ is their centre-centre separation.  The particles interact via a
Hookean force law, $f_{ij} = k(d-r_{ij})$, for $r_{ij} < d$, and zero
otherwise, {\it i.e.}  a one-sided linear spring. In this study, the particle
stiffness $k$ merely parameterises the force scale and for convenience is set
to unity, along with $d$ and $m$. Distances are reported in units of $d$ and
all other quantities are appropriately non-dimensionalised.

Configurations with different amounts of disorder were generated by adding
defects to the original fcc array. In this study, these defects were
introduced by randomly removing particles from the original array then
allowing the configuration to relax into a different mechanically stable state
at the same $\phi$. Removing more particles prior to the relaxation process
resulted in more disorder. As a result the number of particles varied from
$N=16384$ for the FCC lattice down to $N=10384$ for the most random packings
studied.  Consequently, the size of the simulation cube, $L$, ranged from
$19.4d < L < 22.6d$ The amount of disorder was quantified by the coordination
number $z$ (average number of contacting neighbours), given in Table
\ref{table1}, which varied from 12 for the fcc array, to below 9 for the most
disordered packing.  Other measures of the packings corroborated the gradual
transition from ordered to disordered configurations \cite{leo16,leo18}.  This
configuration-generation protocol provided a suitable manner in which to
control the amount of disorder.
\begin{table}[h]
  \caption{Configurations are labelled by $Ci$ and their coordination number $z$. The \rm{fcc} array has a value $ = 12$.}
\begin{tabular}{c|ccccccc}
  \hline
  $Ci$ & $FCC$  & C2 & C4 & C6 & C8 & C10 & C12 \\ \hline
  $z$ & $12$ & $11.85$ & $11.71$ & $11.22$ & $10.68$ & $10.29$ &  $8.92$ \\ \hline
\end{tabular}
\label{table1}
\end{table}
The packings are labelled accordingly: $C[1-6]$ correspond to quasi-ordered
assemblies, $C[7-11]$ weakly disordered, and $C12$ fully disordered
\cite{leo18}. Table \ref{table1} catalogues the packings used in this work
with their respective coordinations.  These data were obtained after averaging
over five independent realisations.

The major aim of this work is to investigate how packing structure affects
the way localised force perturbations are transmitted through sphere packings.
In an effort to eliminate other contributing factors that may strongly
influence the response properties, a particular geometry is used to study
stress transmission. Specifically, to probe the nature of stress transmission
inside the packings, a localised, isotropic perturbation was imposed within
the central region of the packing by increasing the diameter of the centrally
located particles by an amount $\delta d$, in the range $10^{-4} \le \delta
d/d < 10^{-2}$. These values for $\delta d$ are directly converted into forces
using, $F_{perturb} = k(\delta d/2)$. Although this method of perturbation is
not standard compared to experimental and previous $2D$ numerical studies, it
provides a convenient method to study the response properties of unconfined
packings in the absence of sidewalls and directional external forces such as
gravity.  Moreover, this method also improves the statistical analysis as
discussed below.

\begin{figure}[h]
  \centering
  \includegraphics[width=8cm]{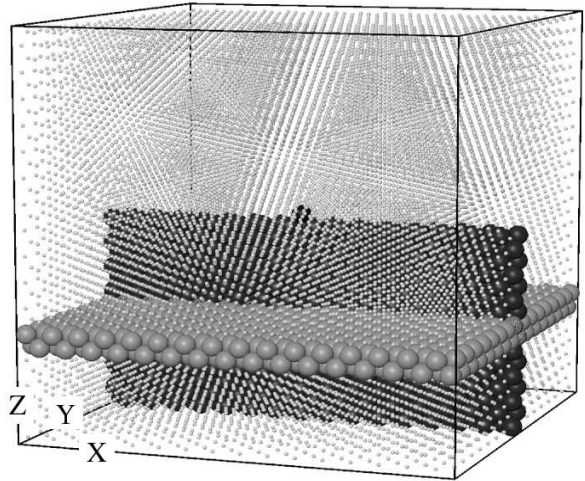}
\caption{Schematic of the simulation configuration. The perturbed region is
  identified by the particle in the middle of the packing.  The planes of
  regular sized spheres represent examples of the visualisation planes for the
  stress transmission presented in the following figures. All other spheres
  are drawn at a reduced size for clarity. Axes are labelled although they are
  all equivalent in this geometry. The simulation cube has side length
  $L=22.6d$.}
\label{fig1}
\end{figure}

For the results presented here to connect with expected theories and
experimental works, \emph{i.e.} to satisfy linearity, it is necessary to gauge
the range of the magnitude of the perturbative force that can be applied to
the packing. This value was based on the average force per particle, $f_{av}$,
in the static packings prior to the perturbation procedure. These average
forces varied over, $10^{-3}kd < f_{\rm av} < 10^{-1}kd$, from the ordered to
most disordered configurations. Thus, the strength of the perturbations was
varied from below to above the average force in most cases. After the
application of this perturbation, the system was again relaxed at the same
initial $\phi=0.742$. This allowed a direct comparison between the initial
stress state $\sigma^{\rm{i}}$ of the configuration before the perturbation
and the stress state $\sigma^{\rm{f}}$, of the final configuration after. The
contact stresses $\sigma$ were computed for all contacting particles inside a
volume $\Omega$ \cite{comment4}, $\sigma_{\alpha\beta} =
\frac{1}{\Omega}\sum_{ij} r^{\alpha}_{ij}f^{\beta}_{ij}$, where the
$\alpha$,$\beta$ are the cartesian components of particles $i$ and $j$,
separated by $r_{ij}$. The effect of the imposed perturbation was computed as
the change in the normal stress between the final and initial states, $\sigma
= \sigma^{\rm{f}} - \sigma^{\rm{i}}$. To reduce statistical uncertainty, for
each independent realisation, the stress was averaged over the equivalent
faces of the periodic cell, {\it i.e.}  $\sigma = (\sigma_{xx}^{+} +
\sigma_{xx}^{-} + \sigma_{yy}^{+} + \sigma_{yy}^{-} + \sigma_{zz}^{+} +
\sigma_{zz}^{-})/6$, where the $+$ and $-$ represent the normal stresses
calculated in the positive and negative direction relative to the position of
the perturbation source. See Fig.~\ref{fig1}.

Figure \ref{fig1} presents a schematic of the simulation geometry. The shaded
regions represent the planes over which the stresses were computed and
visualised in the following figures. The dark gray plane is denoted the
direction parallel to the direction of stress transmission, whilst the light
gray plane is perpendicular to the transmission direction.  The stress
response in the plane above the ($+z$) and below ($-z$) the point of
perturbation are equivalent because, a) the packings are isotropically
compressed, b) there is no gravity acting on the system ({\it i.e.}
directional external forces), and c) because the perturbation itself is
isotropic. Likewise, the $+x$ and $-x$ and $+y$ and $-y$ are also all
equivalent. Observing the stresses in these planes allows a direct comparison
to $2D$ results. This procedure might appear somewhat artificial, however,
there is a close resemblance between this simulation protocol and previous
experimental studies. One must simply imagine taking the simulation cube and
placing any of its faces on a solid floor and then measuring the pressure
profile at this boundary. The benefit of this particular simulation scheme is
the ease with which one can gain `depth' dependent information, where `depth'
is the mean distance from the source of the localised perturbation. Therefore,
rather than generating different packings at different heights, one can simply
look at different slices within the packing \cite{comment9}. Moreover, the
protocol implemented here could be realised in experiments on colloidal
glasses and foams.

\section{Results}
\begin{figure}[h]
\centering
\includegraphics[width=2.5cm]{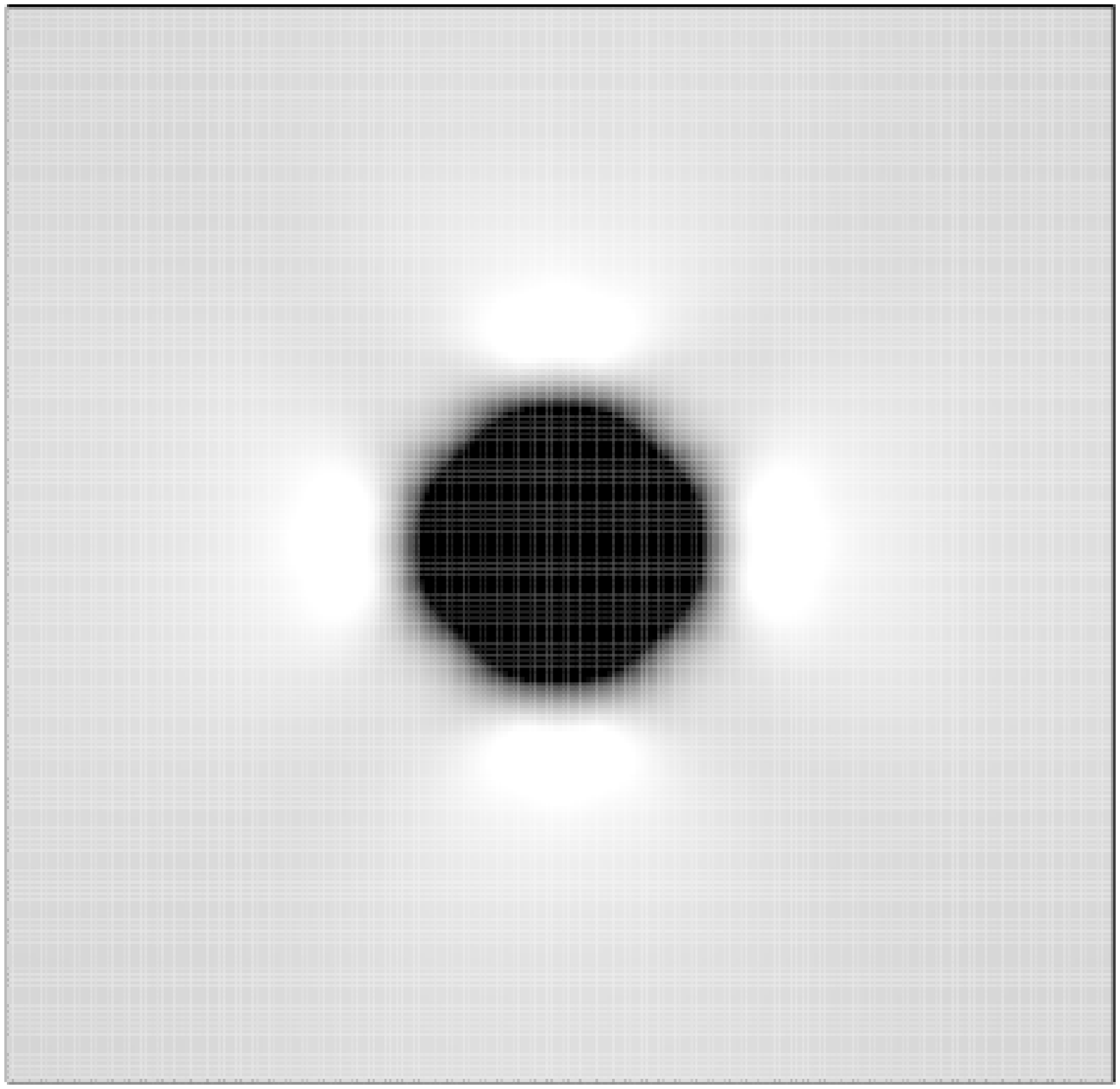}\hfil
\includegraphics[width=2.5cm]{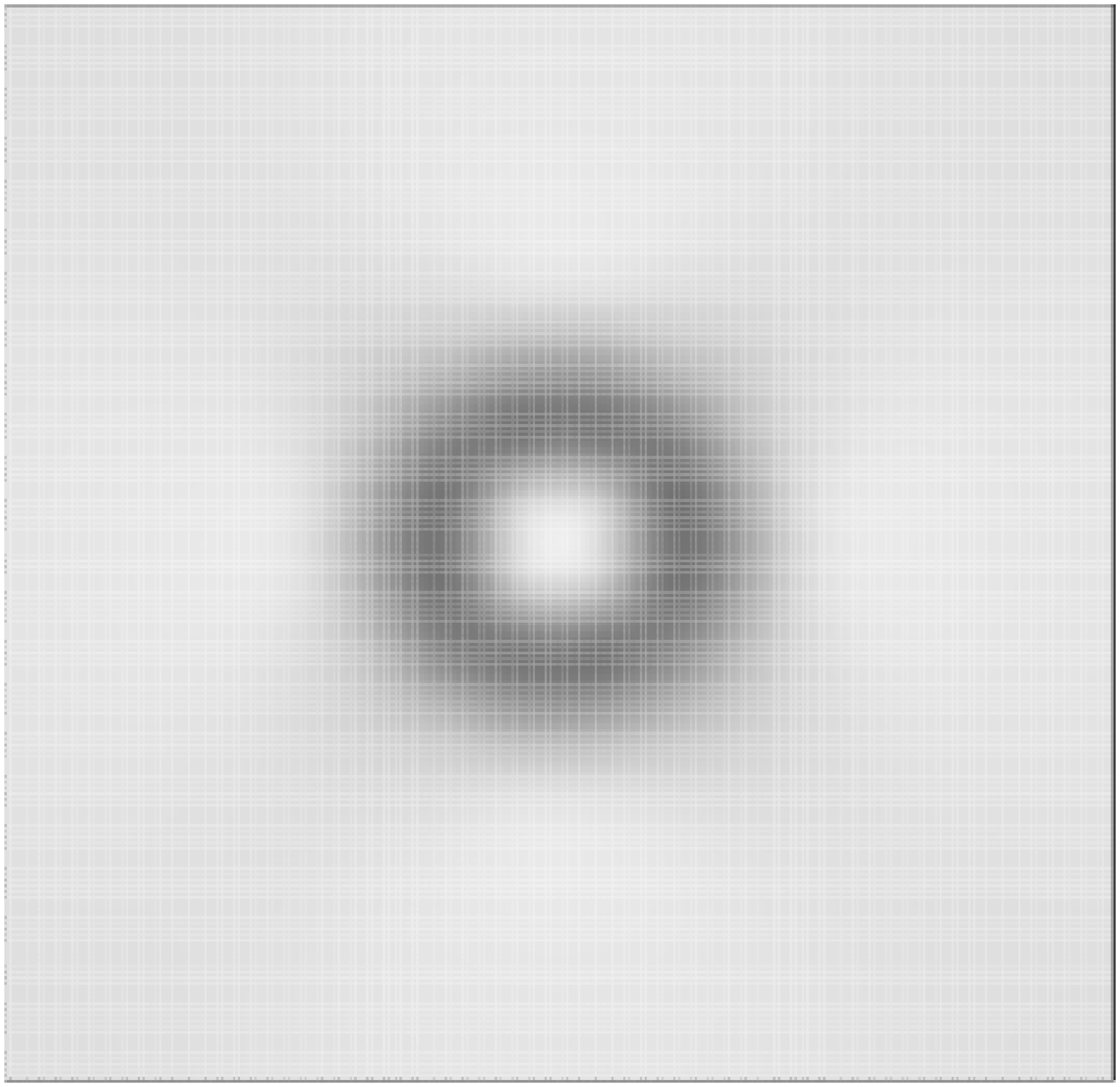}\hfil
\includegraphics[width=2.5cm]{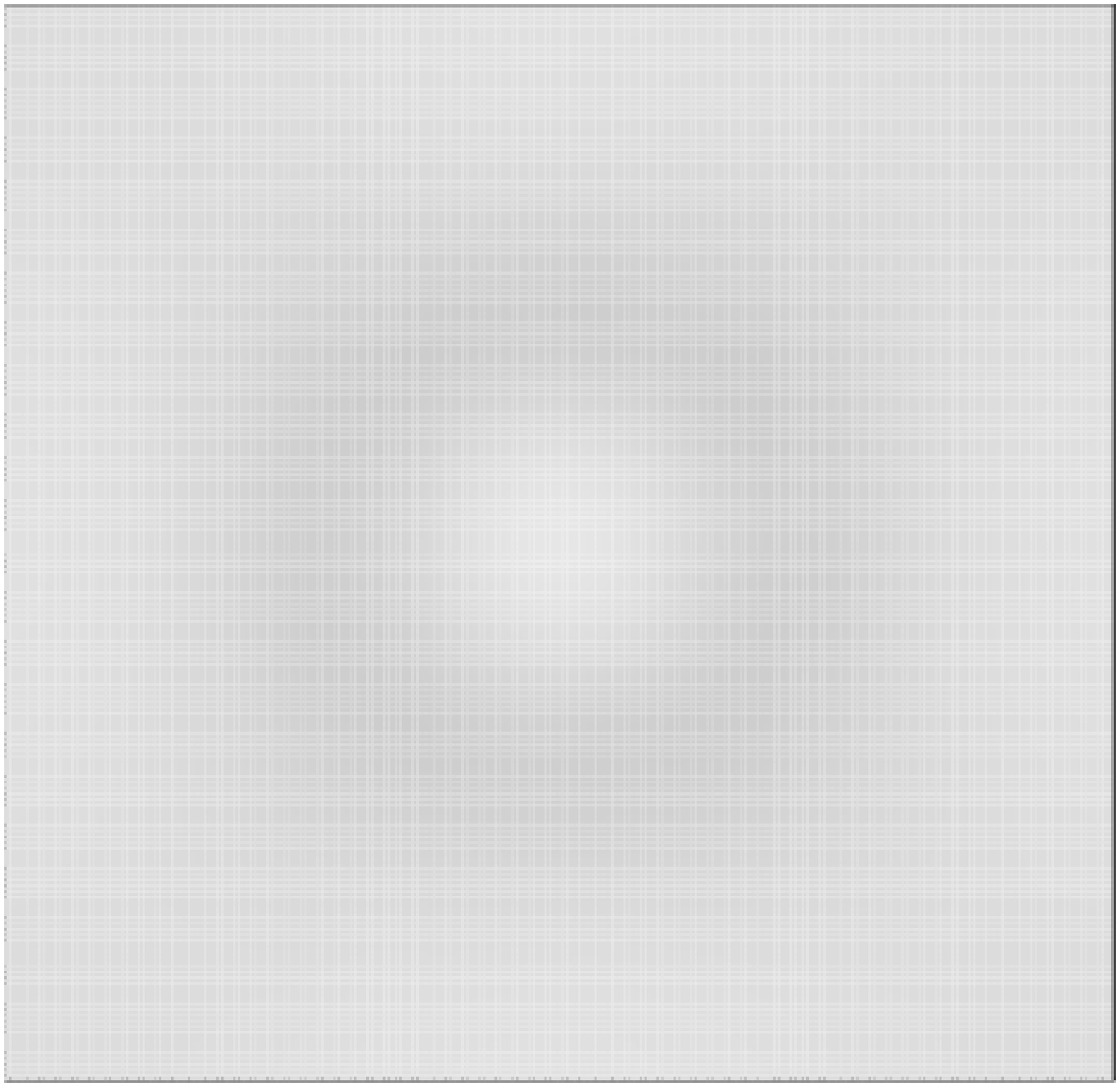}\hfil\\
\bigskip
\includegraphics[width=4cm,height=8cm,angle=-90]{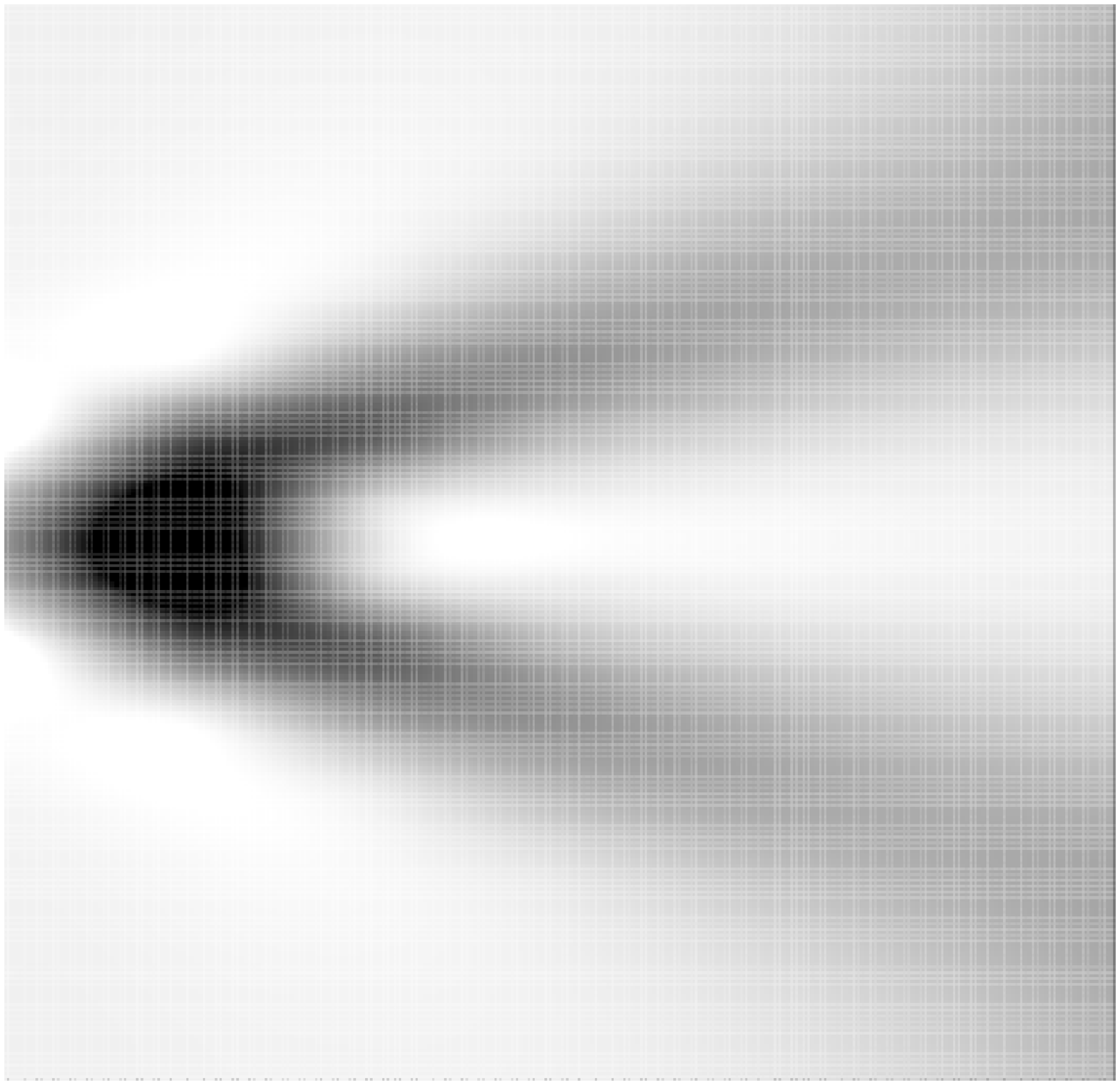}\hfil
\caption{Stress response inside a weakly disordered packing, $C2$, due to
  perturbative force $\delta d = 0.0001$. Top panel: Stress maps of
  approximately $\sim 22.6d \times 22.6d$ in size, corresponding to three
  planes parallel to the light gray plane in Fig.~\ref{fig1}, respectively
  located at $h=2\sqrt{2}$, $3\sqrt{2}$, and $5\sqrt{2}$, from left to right.
  There is a qualitative change in the nature of the response with increasing
  distance from the perturbation source. The strength of the response also
  decays with distance from the source. Bottom: Stress map for a central plane
  corresponding to the dark gray plane in Fig.~\ref{fig1}, with dimensions
  $\sim 22.6d$ wide along the $x$ axis and extending out $\sim 12d$ in the $z$
  direction away from the perturbation source. Darker shades indicate larger
  $\sigma$.  Intensity scale same in all panels.}
\label{fig2}
\end{figure}

Figure \ref{fig2} shows stress response maps for the weakly disordered packing
C2, for one value of the perturbation force. The top panels correspond to
different planes perpendicular to the direction of stress transmission (c.f.
light gray plane in Fig.~\ref{fig1}). The different panels of Fig.~\ref{fig2}
show the stress at different `depths' $h$, with $h=0$ defining the central
plane of the packing, coincident with the perturbation source. While the
response function gradually decays with distance from the perturbation, in the
immediate vicinity of the perturbation (top left panel), the response is
localised (single lobe), capturing the finite extent of the imposed
perturbation. This dramatically changes character away from the source of the
perturbation where the response becomes strongly anisotropic (ringed
response). The bottom panel in Fig.~\ref{fig2} shows the plane parallel to the
direction of stress transmission (c.f.  dark gray plane in Fig.~\ref{fig1})
for the same C2 packing. The response function is directed along
characteristic, broadening pathways, leaving the region between these
high-stress paths experiencing a relative depletion of stress. This view is
similar to the picture put forward in Ref.~\cite{nagel6} on stress
transmission through ordered arrays of glass beads.

\begin{figure}[h]
  \centering
  \includegraphics[width=7.5cm]{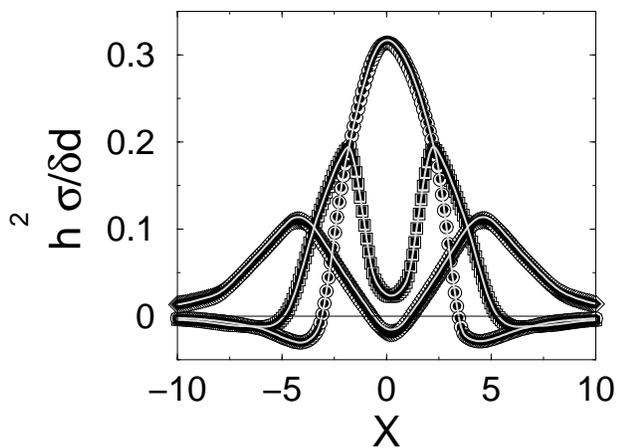}
  \caption{Stress profiles rescaled by the `depth', $h$, from perturbation
    source and the magnitude of the force, $\delta d$, taken from central cuts
    across the planes shown in the top panels of Fig.~\ref{fig2} (packing
    $C2$). The $x$-axis denotes the distance across the packing face. Each
    trio of data are for $h = 2\sqrt{2}$ (circles), $3\sqrt{2}$ (squares), and
    $5\sqrt{2}$ (diamonds), for different values of the perturbing force:
    $\delta d =$ 0.0001 (open), 0.001 (filled), and 0.01 (grey line).}
\label{fig3}
\end{figure}

To further quantify the nature of the response, and in an effort to compare
with equivalent $2D$ results, the stress profiles shown in Fig.~\ref{fig3} are
central cuts taken from (\emph{e.g.}, the top panels in Fig.~\ref{fig2})
slices at different $h$ (equivalently, \emph{e.g.} bottom panel in
Fig.~\ref{fig2}) inside the C2 packing for different perturbation magnitudes
$\delta d$. This clearly demonstrates that close to the source the response is
{\it single-peaked}, crossing over to a {\it double-peaked} response further
away. Not only are the rescaled stresses independent of $\delta d$, but no
contacts were broken in the perturbation process, thus confirming the linear
response regime. To check the robustness of these results, qualitatively
similar data were obtained when $k$ was varied over 5 orders of magnitude, and
the degree of over-compression, $\phi-\phi_{\rm fcc}$, covered 3 orders of
magnitude (results not shown here).

\begin{figure}[h]
  \centering
  \includegraphics[width=8cm]{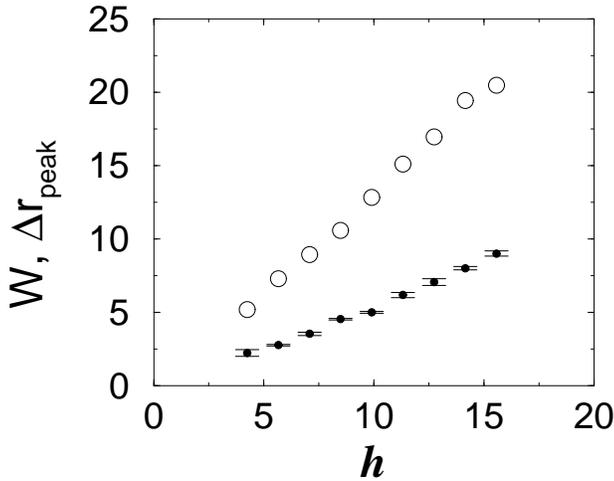}
  \caption{Stress peak analysis for configuration C2 that exhibits a
    double-peak response function. The width $W$ ($\bullet$), measured
    at half height of the stress response peak averaged over the two
    peaks, and the separation between the two peaks, $\Delta r_{\rm
      peak}$ ($\circ$).}
\label{fig4}
\end{figure}

The results of Figs.~\ref{fig2} and \ref{fig3} indicate that away from
the source of the perturbation, the response functions point toward
possible agreement with the hyperbolic wave theory of force
propagation \cite{cates3}, which predicts a double-peaked response
function with peak widths, $W$, scaling diffusively with depth, $W
\sim \sqrt{h}$. To determine which class of theory these results
belong to, Fig.~\ref{fig4} shows the depth dependence of $W$ and the
peak separations $\Delta r_{\rm{peak}}$. From Fig.~\ref{fig4} the
characteristic features of the double-peaked response function are
seen to vary linearly with depth, $W \sim h$, and therefore do not
belong to the hyperbolic class of solutions. This suggests that the
response properties of the weakly disordered, frictionless arrays
studied here are likely to be described by three dimensional
anisotropic elasticity theory.

Although a developed formalism for anisotropic elasticity exists, there are no
analytic solutions with which to directly compare these simulation results.
Alternatively, an empirical fitting form can be used to describe the data.
Motivated by twin-peak, hyperbolic models, Breton {\em et al.} \cite{claudin3}
introduced the Convection-Diffusion (CD) equation that mimics the response
properties of a $2D$ lattice. The CD equation is characterised by a
propagation, or `speed', coefficient $c$, that determines the direction of the
stress response, and a `diffusion coefficient' $D$, characterising the
diffusive spreading of the two peaks with depth. In its original form, the
resulting scalings have the width of the peaks, $W = \sqrt{2Dh}$, consistent
with a wave-like hyperbolic picture. This was modified into the {\it
  Convection-Wave} (CW) equation by Geng {\em et al.}  \cite{behringer5}, in
an effort to deduce whether their experimental data was better described by
the hyperbolic approach or more in line with an elliptic model. To that end,
the CW equation is extended here for $3D$ systems,
\begin{equation}
  \sigma(h) = \frac{F}{2(2\pi)^{3/2}\omega^{2} h^{2}}\left(e^{-\Delta_{-}^{2}/2\omega^{2}h^{2}} + e^{-\Delta_{+}^{2}/2\omega^{2}h^{2}} \right).
\label{eqn2}
\end{equation}
$F$ is the magnitude of the perturbing force, $r$ is the in-plane
radial distance from the central axis of the packing, and
$\Delta_{\pm} = (r \pm c'h)$. Equation \ref{eqn2} guarantees a twin
peak response with peak widths scaling linearly with depth, $W=\omega
h$, where $\omega$ and $c'$ depend on the packing structure.

\begin{figure}[h]
  \centering
  \includegraphics[width=8cm]{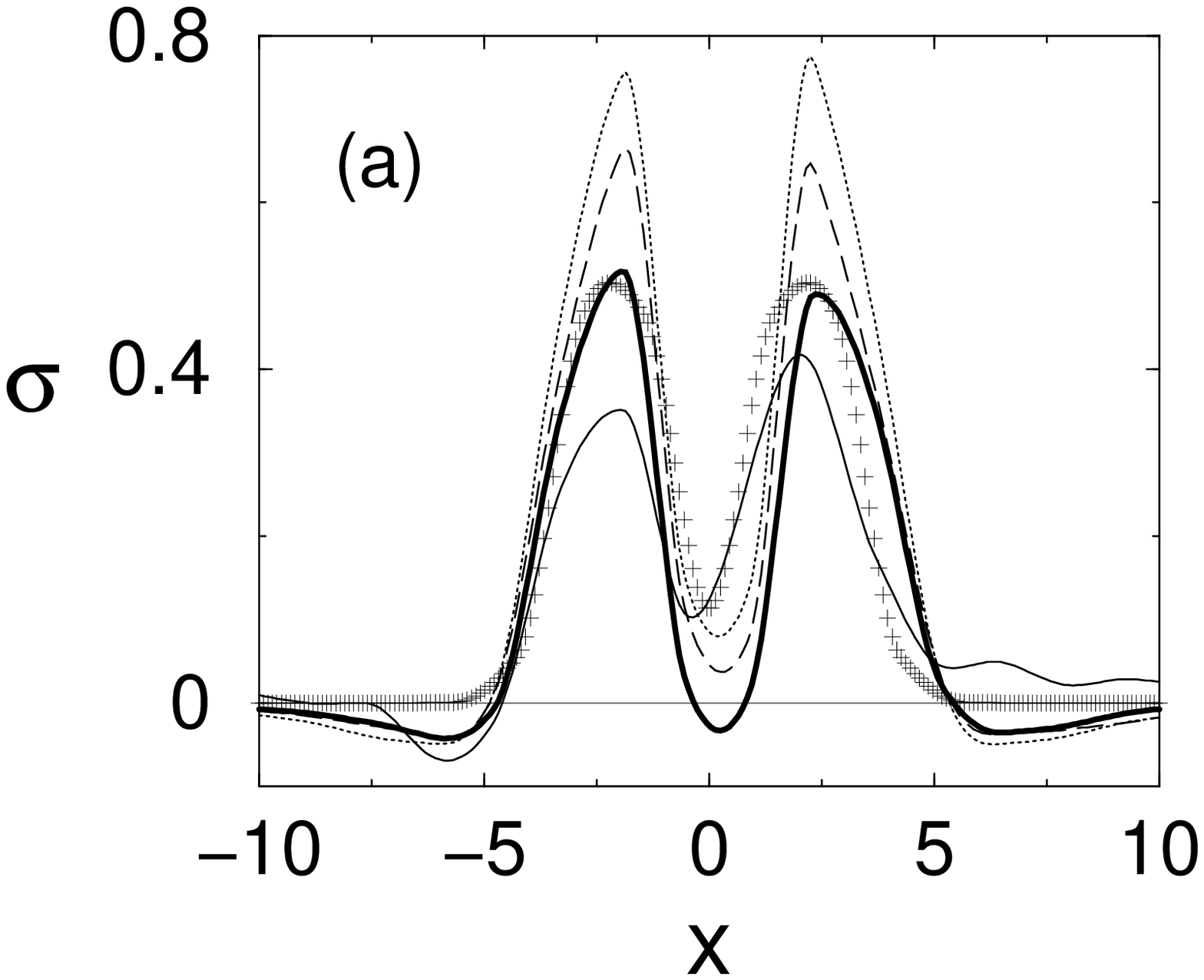}
  \includegraphics[width=8cm]{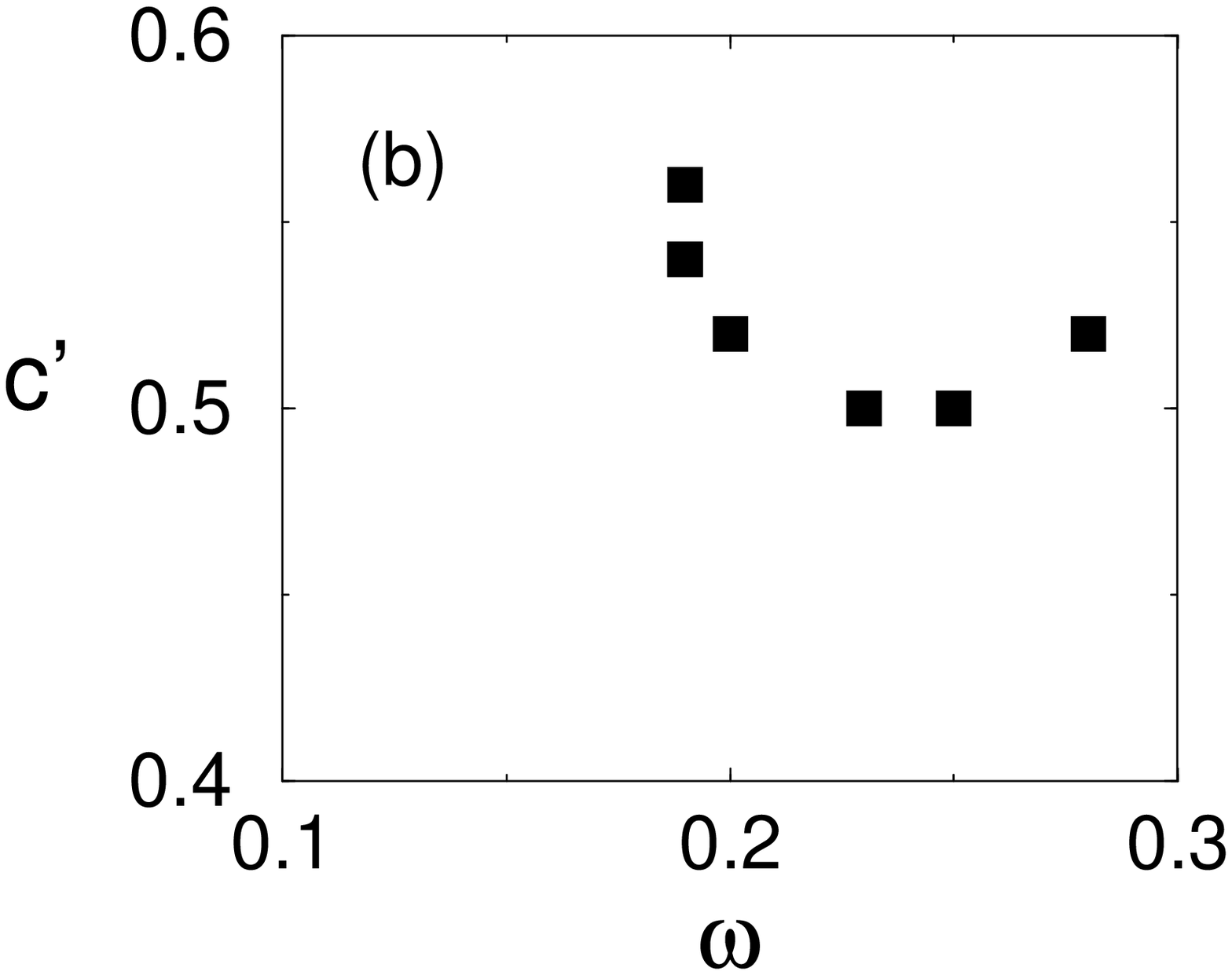}
  \caption{(a) Scaled stress profiles at $h=3\sqrt{2}$ for $\delta=0.01$, in
    the weakly disordered regime. Peak heights decrease with increasing
    disorder from highest, C2, to lowest, C8.  The symbols (+) are a fit of
    Eq.~\ref{eqn2} to the data for C6 (thick solid line), with best-fit
    parameters for $c'$ and $\omega$, and is representative of the goodness of
    the fit for the other configurations. (b) The values of the fitting
    parameters $c'$ and $\omega$ used in Eq.~\ref{eqn2} extracted from fitting
    to the packings $C2-C8$.}
  \label{fig5}
\end{figure}

Figure \ref{fig5}(a) shows the stress response profiles at one depth for a
given perturbation, over a range of weak disorder. The double-peak profiles
evolve with disorder: the amplitude of the response decreases with increasing
disorder, although over this range in disorder, there appears to be little
change in the location of the peaks. Equation \ref{eqn2} captures the main
features of the profiles - the characteristic double-peak - yet it fails to
accurately describe the full response, such as the dip in the profiles either
side of the peaks. A representative fit of Eq.~\ref{eqn2} (symbols) to the C6
configuration (thick solid line) is shown. A criterion was used in the fitting
procedure that attempted to match the peak positions as closely as possible
between the simulation data and Eq.\ref{eqn2}. Despite these minor
inadequacies, Eq.~\ref{eqn2} can be used to characterise the profiles in the
weakly disordered regime through the coefficients $c'$ and $\omega$, which are
obtained as fitting parameters. Over the range of disorder explored in
Fig.~\ref{fig5}(a), the values of these fitting parameters are shown in
Fig.~\ref{fig5}(b). The upper left points correspond to the strongly ordered
systems whereas the lower right points correspond to the weakly disordered
configurations. These coefficients not only follow the same trend, but are
somewhat similar, to those obtained in ordered and weakly-disordered $2D$
experiments \cite{behringer5}.

\begin{figure}[h]
\centering
  \includegraphics[width=2.5cm,height=2.5cm]{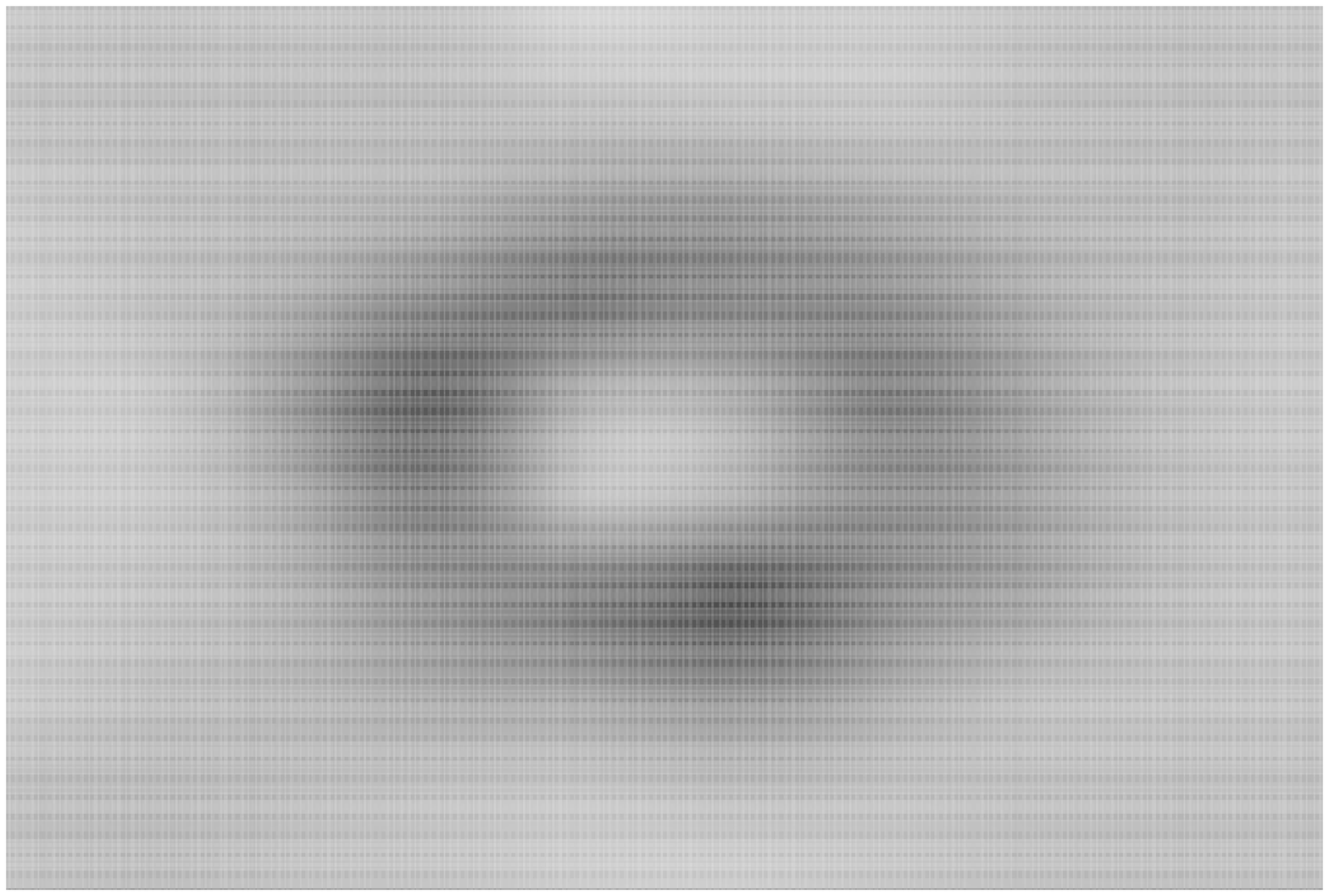}\hfil
  \includegraphics[width=2.5cm,height=2.5cm]{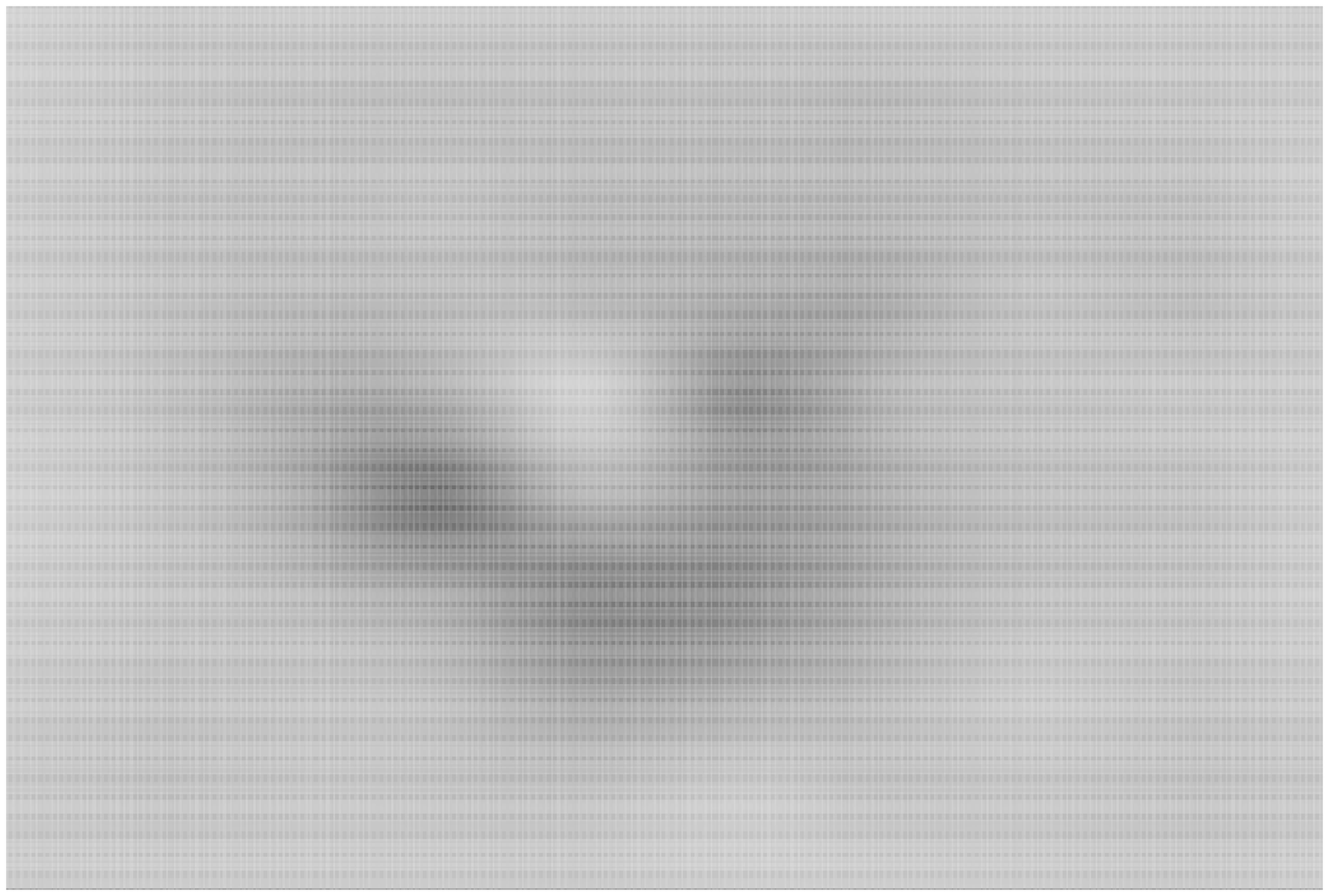}\hfil
  \includegraphics[width=2.5cm,height=2.5cm]{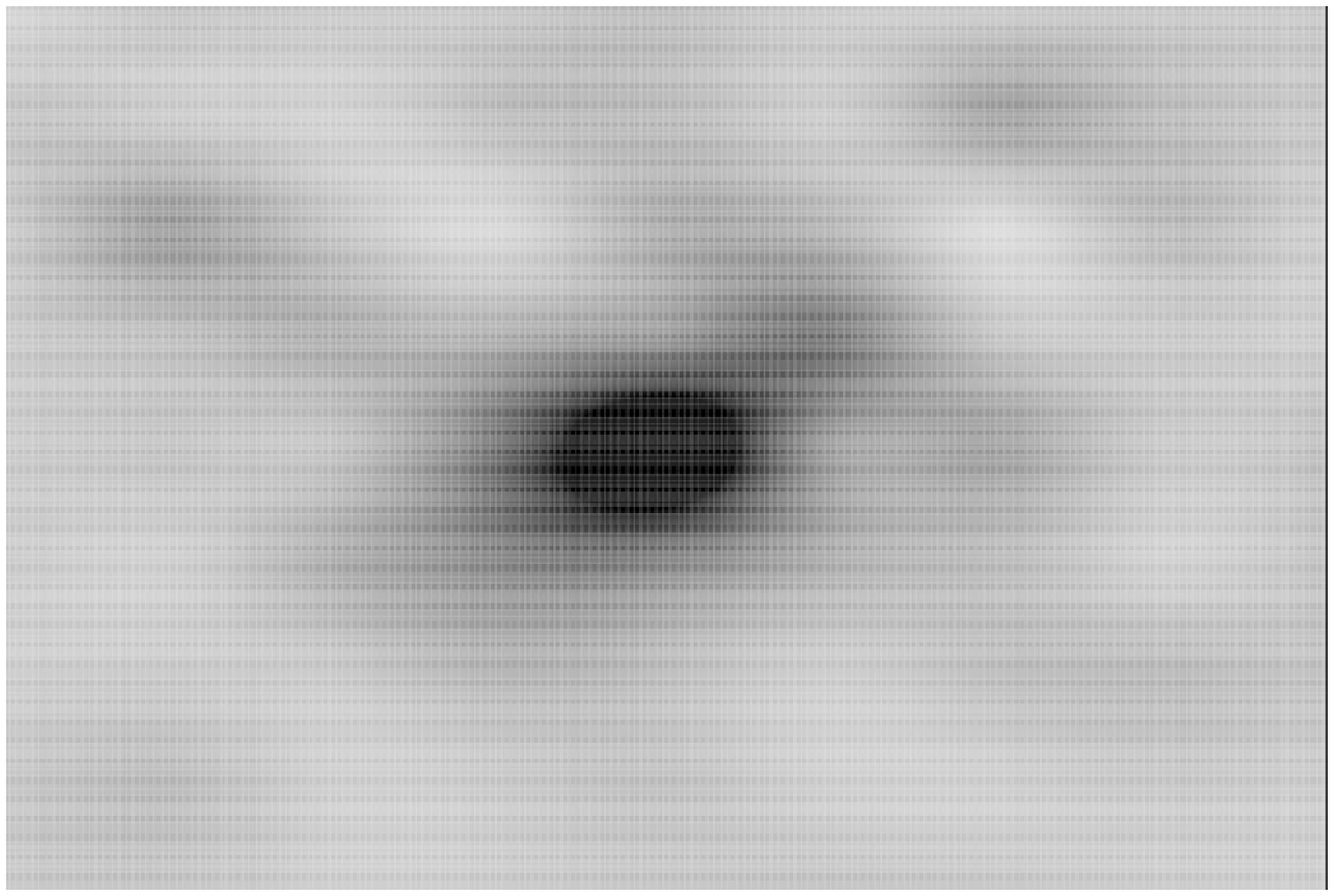}\hfil
  \caption{Stress response at $h=3\sqrt{2}$ for $\delta d = 0.001$, inside
    packings with increasing disorder from left to right: $C8$, $C11$, and
    $C12$. Size of slice $\sim 20d \times 20d$. Shading is the same in all
    panels.}
\label{fig6}
\end{figure}

\begin{figure}[h]
\centering
  \includegraphics[width=8cm]{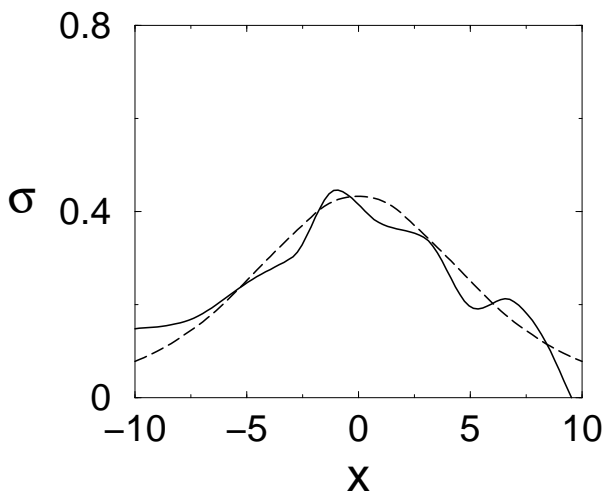}
  \caption{Comparison between the Boussinesq theory, Eq.~\ref{eqn1} (dashed
    line) and the stress response for the most disordered configuration C12,
    which resembles an overcompressed random close packing. At $h=3\sqrt{2}$
    inside the packing for $\delta d = 0.001$.}
\label{fig7}
\end{figure}

When further disorder is introduced, the anisotropic response profiles become
less well-defined and there is a crossover to a more isotropic response of the
single-peak variety. This is illustrated in the stress maps shown in
Fig.~\ref{fig6}, for configurations with moderate-large disorder. In an effort
to improve averaging over the growing stress fluctuations, both the size of
the perturbed region and the averaging volume were varied. It is expected that
even for frictionless particles, disorder promotes an isotropic, elastic-like
response \cite{goldenberg3}. For the most disordered system (C12), whose
structure resembles that of an overcompressed random closed packed state, a
single-peaked response can be clearly resolved. This is better illustrated in
the profile shown in Fig.~\ref{fig7}. A comparison to the Boussinesq theory of
Eq.~\ref{eqn1} shows reasonable agreement \cite{comment8}. Further promoting
the idea that isotropic random packings are more likely to exhibit isotropic
elastic-like stress behaviour.

\section{Discussion}
The procedure described here presents only one of a number of avenues through
which the mechanical response of a material may be investigated.
Traditionally, shear tests have been used to determine constitutive parameters
through which material, or elastic, constants can be obtained from linear
stress-strain relations. Moreover, for the explicit study of response
properties of the type studied here, the elastic constants must be initially
determined which can then be input into finite element analyses routines
\cite{claudin5}. However, in terms of computational time such procedures are
just as costly, if not more so, as the perturbation protocol implemented here.

To reiterate, the goal here was to make a connection with experimental studies
on similar systems. The results presented here are the first in a series of
investigations on the mechanical response in particulate media. In this
initial study it has been shown that the force response method provides a
quantitative description of the elastic properties of particle packings
consistent with previous studies. To that end, the protocol implemented here
has been verified as a suitable procedure through which we can investigate
mechanical response problems. Even though the particular geometry used is not
conventional from an experimental standpoint it does offer a method to
determine the influence of different parameters on the mechanical behaviour of
sphere packings. The focus here was \emph{structure} and is likely relevant
not only to compressed elastic sphere-packings, but also foam dispersions.

\section{Conclusions}
For the first time, a systematic study of the nature of stress transmission in
response to localised force perturbations has been obtained \emph{inside}
three dimensional packings of frictionless spheres using discrete element
simulations, in a particular geometry that avoids the influence of walls. The
manner in which stresses are transmitted crucially depends on the distance
from the source of the perturbation and the structure of the packing. For
small deviations from the fcc structure, the response in the vicinity of the
perturbation source retains the localised form of the perturbation, and
appears as a single-peak response function. This crosses over to an
anisotropic response function of the double-peak variety further from the
source. This double-peak feature is suitably captured by the modified
Convection-Wave equation; peak widths scaling linearly with distance, rather
than the diffusive form of this equation \cite{claudin3}. These results are
consistent with a generalisation from $2D$ to $3D$ anisotropic elasticity
theories \cite{claudin4}. Further disorder deforms the double-peak features
associated with the anisotropic response function until there is a crossover
to a single-peak response for randomly packed spheres consistent with an
isotropic elastic description. Moreover, this crossover is qualitatively
matched by the Boussinesq equation, Eq.~\ref{eqn1}, which again indicates that
the procedure implemented here probes mechanical properties that is simple and
quite accurate \cite{goldenberg6}. This will be particularly useful in future
studies on random packings with different coordination numbers with and
without friction.

It is known experimentally the stress response of a disordered, frictional
granular packing qualitatively agrees with isotropic elasticity theory
\cite{clement1}. Simulations of two dimensional systems also point towards
agreement with isotropic elastic models \cite{goldenberg3}. Studies of three
dimensional frictional ordered and disordered packings and gravity packings
remains work in progress.

\begin{acknowledgements}

  I thank P. Claudin and J. Socolar for fruitful discussions, and one of the
  referees for their detailed and supportive critique. I gratefully
  acknowledge the support of an ORDA Faculty Seed Grant at SIU and the
  National Science Foundation under Grant No. CBET-0828359.

\end{acknowledgements}


\begin{thebibliography}{44}
\bibitem[{comment4({\natexlab{a}})}]{comment4}
 For the weakly disordered systems, the results are
  insensitive to the choice of $\Omega$, provided it is larger than the
  interparticle spacing. In that case $\Omega = \sqrt{2}$. For more disordered
  configurations, $\Omega = \frac{3}{2}\sqrt{2}$.
\bibitem[{comment9({\natexlab{b}})}]{comment9}
 The results presented here are qualitatively similar to
  packings in a gravitational field resting on a solid boundary.
\bibitem[{comment8({\natexlab{c}})}]{comment8}
 A fitting of the data to the response expected within an
  infinite elastic medium was also attempted, with the Poisson ratio of the
  material now entering as a free parameter \cite{landau2,timoshenko1}. This
  excercise did not shed any further light on the analysis.
\end{thebibliography}
\end{document}